# Intrinsic switching leads to oxygen diffusion and breakdown in hafnia ferroelectrics

**Authors:** Xabier Diaz de Cerio[1]*, Iñigo Robredo-Magro[1], Jorge Íñiguez-González[1,2]*

**Affiliations:**

[1]Smart Materials Unit, Luxembourg Institute of Science and Technology (LIST); Esch-sur-Alzette, Luxembourg.

[2]Department of Physics and Materials Science, University of Luxembourg; Belvaux, Luxembourg.

*Corresponding author. Email: xabier.diaz-de-cerio-palacio@list.lu (X.D.C); jorge.iniguez@list.lu (J.Í.-G.)

**Abstract:** Conventional ferroelectrics exhibit well-defined polarization states linked through electric switching. In fluorite-structured ferroelectrics like hafnia, though, switching and oxygen diffusion seem to coexist, enriching the nature of ferroelectricity. Here we address the intrinsic, room-temperature switching and diffusion kinetics of hafnia ferroelectrics using machine-learning molecular dynamics. We identify two distinct switching mechanisms that are both active at realistic time scales. Critically, our simulations reveal that, because the lattice does not dissipate fast enough the heat originating from localized switching events, these two processes concatenate in an avalanche-like manner leading to oxygen conduction. Our results thus show that intrinsic switching leads to breakdown in hafnia ferroelectrics. They also suggest how this outcome might be avoided through suitably designed field pulses.

**Main Text:** Hafnia ($HfO_2$) is one of the most promising ferroelectrics due to its compatibility with CMOS technology and robustness against device downscaling (*1–5*), which set it apart from conventional perovskite ferroelectrics. However, the remarkable stability of ferroelectricity in hafnia thin films comes together with very high coercive fields needed to switch polarization, their proximity to the breakdown threshold representing a crucial limitation in the development of reliable devices (*6–8*).

The origin of this drawback is still obscured by the limited understanding about the nature of ferroelectricity in this material. The mechanisms enabling the stabilization of the ferroelectric phase and the switching of polarization have been addressed using traditional pseudosymmetry arguments and lattice mode analysis based on first-principles calculations at 0 K (*9–15*). In contrast to conventional materials –e.g. perovskite ferroelectrics–, multiple switching paths and reference centrosymmetric structures have been identified in hafnia. In particular, the same state can exhibit switching leading to polarization changes of opposite sign (*13*, *16*, *17*). In addition, all the proposed paths involve large energy barriers and ionic displacements in the order of a lattice constant, which suggests that switching in hafnia can be understood as a highly energetic bond-breaking process of a topotactic nature (*18*). This idea has been further strengthened by recent experimental evidence suggesting the coexistence of oxygen (O) diffusion and ferroelectric switching, which implies that measured changes of polarization are likely affected by considerable ionic currents (*19–21*).

This unconventional scenario questions the suitability of the traditional perturbative and pseudosymmetry assumptions, which are the basis of Landau models, for understanding hafnia-based ferroelectrics at room temperature. Indeed, the large coercive fields and ionic displacements require attention to heat dissipation and collective kinetic effects if one aims to capture the nonequilibrium physics of switching and its potential relation to O diffusion. Furthermore, stochastic processes like those call for a statistical treatment of thermally activated rare events at realistic time and length scales.

Here we address this problem through machine-learning molecular dynamics (MD) simulations of switching and diffusion in ferroelectric hafnia at room temperature considering relevant length and time scales. We focus on intrinsic behaviors, working with defect-free samples and employing periodic boundary conditions. While idealized, this case lends itself to an accurate computational treatment and allows us to build our understanding systematically. In such defect-free conditions we can expect switching and breakdown fields be the most separated, as conduction should be hardest to achieve. Yet, we find that ferroelectric switching and electric breakdown occur concomitantly, even if only intrinsic mechanisms are allowed.

## Results and discussion

We perform room-temperature MD simulations of the ferroelectric orthorhombic oIII ($Pca2_1$) phase of hafnia under external electric field $\mathcal{E}$ employing a machine-learned interatomic potential derived from accurate and extensive first-principles data. As shown in Fig. 1(A), the oIII polymorph is characterized by an alternation along $Y$ of active (red) and inactive (gray) two-dimensional O channels extending in the $XZ$ plane, with the polar axis along $Z$ (*3*, *10*). To identify dominant switching mechanisms and to characterize their rates, we consider system sizes of ~10,000 atoms and timescales of ~10 ns, working in constant temperature conditions. Switching and diffusion cause local temperature fluctuations that significantly affect the atomic dynamics, and a realistic treatment of heat dissipation is key to avoid artifacts in our predictions (*22*). We achieve this by using stochastic boundary conditions (SBC) (*22*, *23*), whereby we set to 300 K the temperature of just the hafnium (Hf) ions at the top and bottom boundaries of the simulation supercell, mimicking the case of a thin film that exchanges heat with the environment through surfaces or interfaces (see Methods and fig. S1 to S5).

We start our study with a simple test to identify the typical magnitude of switching and diffusion fields. Noting that the sign of the polarization of a given polar state is ill-defined (*16*, *24*, *25*), we adopt the following approach. We work with a periodically repeated supercell of 3.13 nm × 7.01 nm × 3.02 nm (6 × 14 × 6 unit-cell repetitions, 6,048 atoms), where the third axis aligns with the polar direction of an oIII-1 monodomain state (Fig. 1(A)). We apply both positive and negative electric fields along the polar axis and identify the switching mechanisms in Fig. 1(B), which shows the response when the strength $|\mathcal{E}|$ increases linearly in time. Beyond the linear dielectric regime, the polarization ($\Delta P_z$) undergoes abrupt changes at threshold fields $\mathcal{E}_t^A = -12.8$ MV/cm and $\mathcal{E}_t^B = 14.6$ MV/cm. Both events, labelled type-A and type-B respectively, correspond to switching from oIII-1 to the state that we denote oIII-2 (Fig. 1(A)). In particular, type-A switching consists of the collective upshift of all O ions, with a swap in their active/inactive character (Fig. 1(C)) (*12*, *21*). Accordingly, the system goes through a *P4₂/nmc* (tetragonal, t)-like transition state (*10*, *12*). By contrast, type-B switching involves the downshift of the active O ions across a Hf plane, while the inactive oxygens are not affected in any significant way (Fig. 1(C)). Here the system passes through a *Pbcm* (oVII)-like state (*11*, *13*, *16*, *26*). We thus find that oIII-1 can be switched under both positive and negative electric field, via two distinct paths that induce polarization changes of opposite sign (*21*).

For the field increase rate employed in Fig. 1(B), we obtain similar threshold fields for type-A and type-B switching. We further assess this in Fig. 1(D), where we apply constant $|\mathcal{E}|$ values below such threshold fields and quantify the time needed to trigger each event ($\tau$) as a function of $1/|\mathcal{E}|$. The solid black (type-A) and red (type-B) lines show that, in both cases, the $\tau$ vs. $1/|\mathcal{E}|$ relation is well captured by the empirical Merz's law,

$$\tau = \tau_0 e^{\mathcal{E}_a/|\mathcal{E}|} \quad (1)$$

where $\tau_0$ is the switching time in the limit of $|\mathcal{E}| \to \infty$ and $\mathcal{E}_a$ is the so-called activation field. Most notably, while type-A switching needs lower $|\mathcal{E}|$ to be activated within the timescales accessible in our simulations ($\lesssim$50 ns), the increase of $\tau$ with $1/|\mathcal{E}|$ is visibly slower for the type-B process. If we assume that the Merz-like behavior in Fig. 1(D) extrapolates to longer timescales (from ~0.1-10 ns in our simulations to ~ms relevant experimentally), this suggests that the type-A and type-B threshold fields will become very close in practice (fig. S6).

This proximity of threshold fields suggests that the application of a coercive field could sustain a continuous flow of oxygen ions via subsequent alternation of type-A and type-B events, such that polarization does not saturate after switching (*21*). Crucially, this would imply that intrinsic switching and breakdown fields are effectively the same, too close to avoid the activation of ionic conduction upon switching in a device.

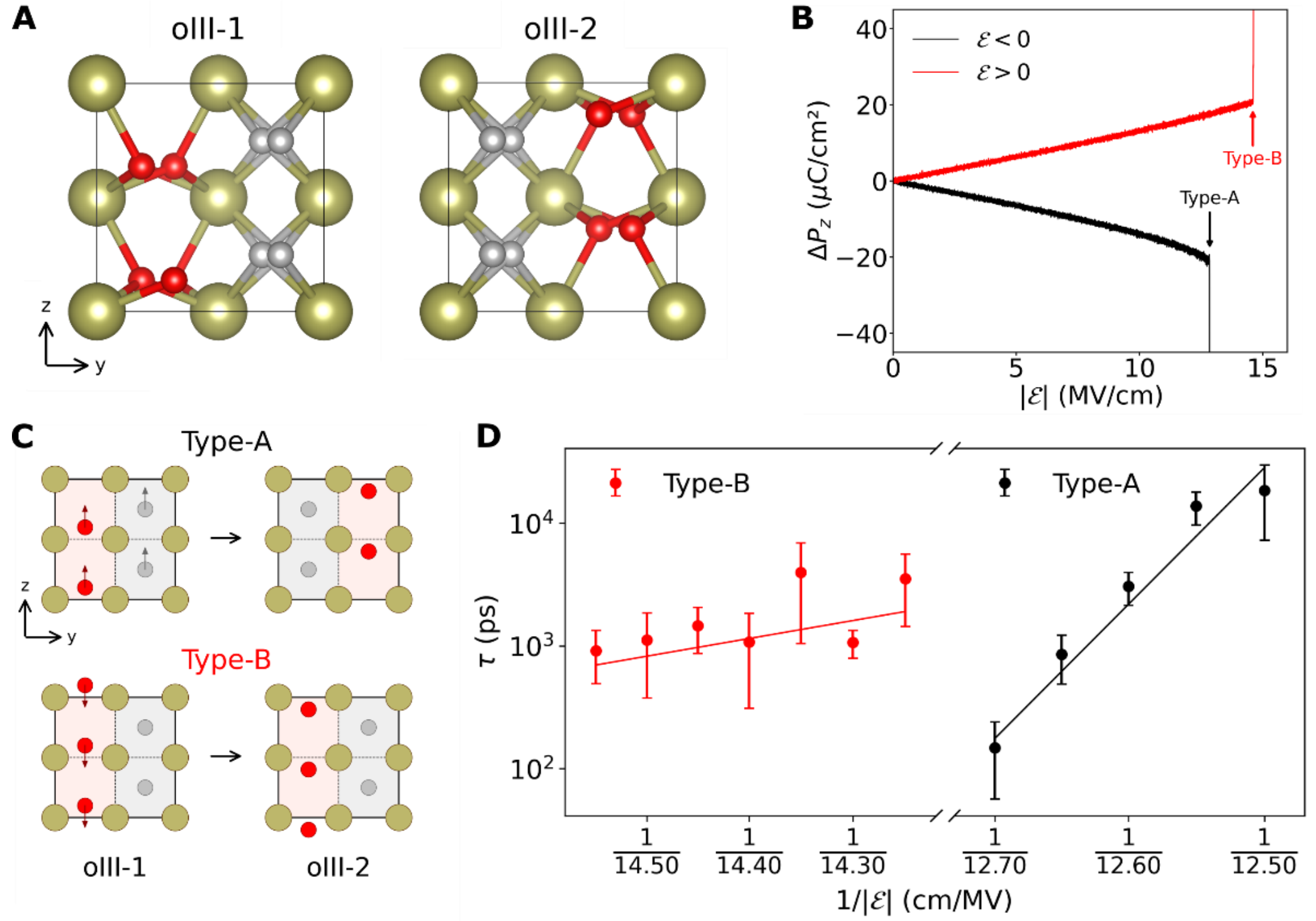

**Fig. 1. Dominant room-temperature switching mechanisms in hafnia.** (**A**) Atomic structure of the oIII-1 (left) and oIII-2 (right) polar states of the oIII (*Pca2₁*) polymorph of hafnia. Active and inactive O atoms are represented in red and grey, respectively. Hf atoms are represented in khaki. (**B**) Induced change in polarization ($\Delta P_z$), calculated with respect to the unperturbed oIII-1 state, as a function of external electric field strength ($|\mathcal{E}|$) for $\mathcal{E} < 0$ (black) and $\mathcal{E} > 0$ (red). $|\mathcal{E}|$ is linearly increased over time at a constant rate of 5 MV/cm per ns. $\mathcal{E} < 0$ and $\mathcal{E} > 0$ induce type-A and type-B switching, respectively. (**C**) Schematic unit-cell representation of type-A (top) and type-B (bottom) switching. (**D**) Switching activation time ($\tau$) as a function of $1/|\mathcal{E}|$ for type-A (black) and type-B (red) mechanisms. The simulation corresponding to each value of $\mathcal{E}$ was repeated 3 times. The averages and standard deviations over the 3 repetitions are represented by the dots and error bars, respectively. The solid lines correspond to the fit of the data to Merz's law (see Eq 1), which yields $\mathcal{E}_a^A = 4033$ MV/cm, for type-A, and $\mathcal{E}_a^B = 694$ MV/cm, for type-B.

We explore this possibility by examining the switching kinetics in a 3.13 nm × 14.02 nm × 3.02 nm cell (6 × 28 × 6 unit-cell repetitions, 12,096 atoms) under constant $\mathcal{E}$. The case where we induce type-B switching first is summarized in Fig. 2 for a representative field of $\mathcal{E}$ =14.5 MV/cm. As shown in Fig. 2(A), the first switching event occurs after 567 ps, with a polarization increase of 4.5 $\mu$C/cm$^2$. To track the nature of this event, Fig. 2(B) shows the time evolution of $P_z$ as a function of *Y* coordinate, obtained as an average over the *XZ*-plane. Remarkably, the observed event corresponds to the type-B switching of a single *XZ*-plane of active oxygens, which leaves the system in a stable configuration formed by one reversed oIII-2 channel embedded in an oIII-1 matrix (bottom half of Fig. 2(B) and movie S1).

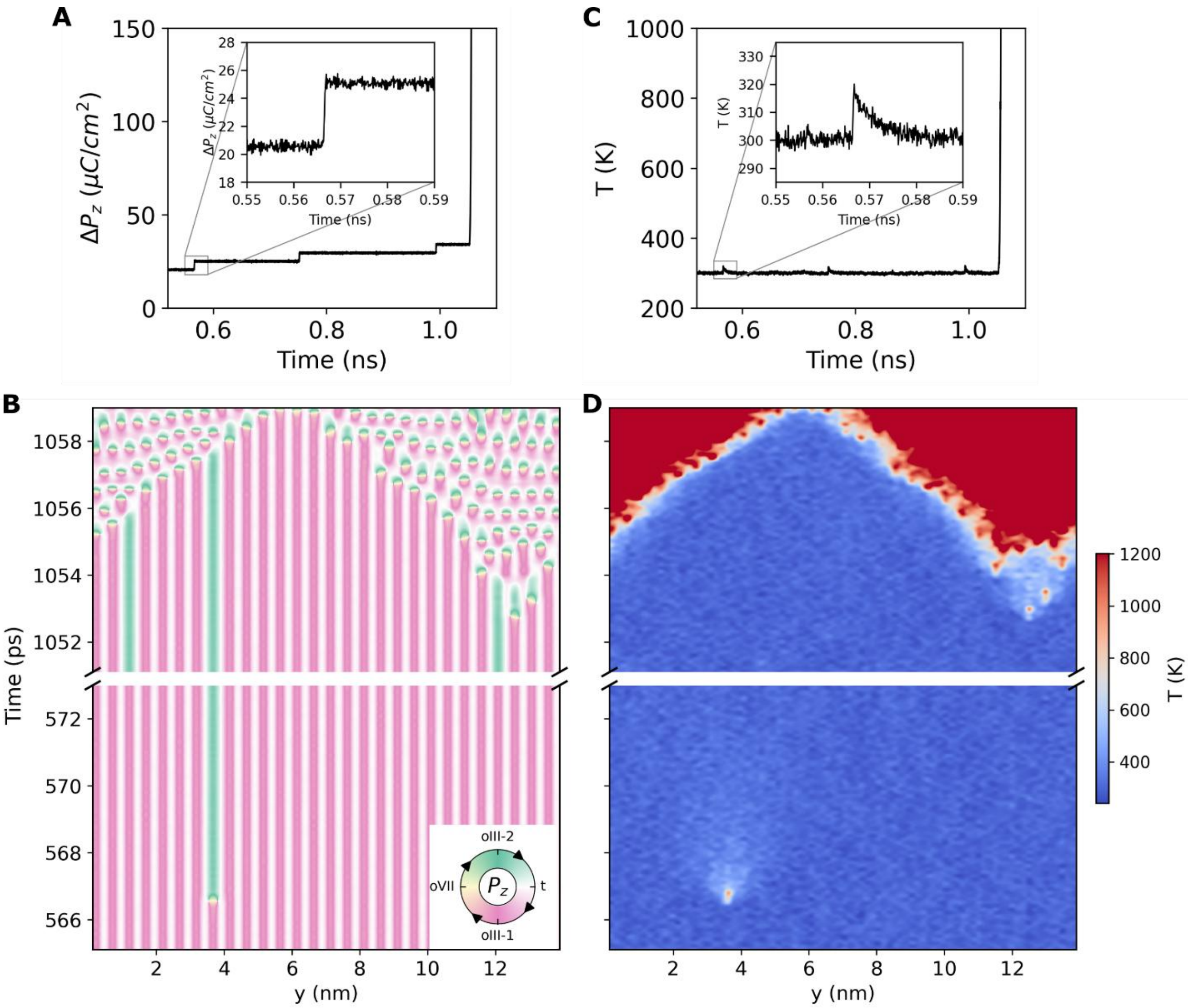

**Fig. 2. Type-B switching kinetics.** Results for an applied field $\mathcal{E} = 14.5$ MV/cm. (**A**) Polarization ($\Delta P_z$) with respect to the unperturbed oIII-1 state as a function of time. Inset: close-up of the first switching event. (**B**) Time evolution of the *Y*-resolved local polarization ($P_z$) averaged over the *XZ* plane. The cell is initially in a monodomain oIII-1 state, with the local configuration of individual O channels alternating between the active oIII-1-like (pink) and the inactive t-like (white) configuration. As shown by the circular colorbar (bottom right corner), the local $P_z$ changes cyclically under a field. The clockwise-pointing arrows in the colorbar indicate the sense of the change of $P_z$ induced by a positive field. Type-B switching from the oIII-1 (pink) to the oIII-2 (green) state passes through an oVII-like (yellow) state. Type-A switching from the oIII-2 (green) to the oIII-1 (pink) state passes through a t-like (white) state. (**C**) Global temperature (T) as a function of time. Inset: close-up of the first switching event. (**D**) Time evolution of the *Y*-resolved temperature averaged over the *XZ* plane.

We get further insights by monitoring globally and locally the temperature of the system. Figure 2(C) shows that switching produces a sudden increase of ≈20K in the global temperature. Figure 2(D) reveals that such fluctuation is initially confined to the switched channel, where it appears as a strong, isolated spike. This localized heat then dissipates into the surrounding channels, as evidenced by the flattening and lateral expansion of the initial spike. Eventually, as the extra heat dissipates through contact with the environment (simulated using SBC), the global temperature decays back to 300K (Fig. 2(C)).

The first switching event is followed by other equivalent type-B events. As shown by the top half of Fig. 2(B) (see $t \leq 1052$ ps), these appear randomly at positions non-adjacent to previously switched channels, resulting in the formation of stable multidomain states. Eventually, at $t \approx 1053$ ps in our simulation, Figs. 2(A,C) show giant increases in $\Delta P_z$ and global temperature. Here, an additional type-B event is immediately followed by a concatenation of type-B and type-A events (top half of Fig. 2(B) and movie S2). This happens initially in adjacent channels, but gradually propagates to the whole supercell in an avalanche-like process that leads to permanent O diffusion. This avalanche can be understood in terms of collective kinetic effects. The local temperature in the top half of Fig. 2(D) shows that the initial event produces a strong localized increase analogous to the one described in bottom half of Fig. 2(D). However, this time the local temperature increase is enough to thermally activate another event in an adjacent channel, which produces a further increase of the local temperature and induces other nearby events. The system thus undergoes what is called a *thermal runaway chain reaction* (*27*, *28*), where every new event increases the probability that additional events occur. This phenomenon leads to the breakdown of the system, characterized by a continuous ionic current and a giant increase in the global temperature.

We find a similar situation when we induce an initial type-A switching, as summarized in Fig. 3 for the representative case of $\mathcal{E} = -12.55$ MV/cm. Here, however, the giant $\Delta P_z$ and temperature increases happen as soon as the first switching event is triggered (Figs. 3(A,B)). Figure 3(C) shows that type-A switching starts with the nucleation of a small oIII-2 domain. The simultaneous upshift of all O ions in a switching nucleus localized around $Y$ = 1.2 nm, and the swap in their active/inactive character, result in the formation of *$P4_2/nmc$*-like domain walls (DWs) featuring adjacent inactive channels. Switching is then completed through the fast lateral growth of the oIII-2 domain. However, before the type-A switching to oIII-2 extends to the whole volume, new oIII-1 domains emerge again through local type-B switching in the transformed oIII-2 region. This is followed by an avalanche of alternating type-B and type-A events, leading to permanent O diffusion (see also movie S3). The local temperature in Fig. 3(D) shows that heating follows the lateral migration of DWs, as the temperature of each channel increases right after it switches. Consequently, the whole simulated system gets increasingly heated, ultimately resulting in thermal runaway. Hence, regardless of whether the field applied to the oIII-1 monodomain state is positive (Fig. 2) or negative (Fig. 3), we ultimately observe that the system breaks down.

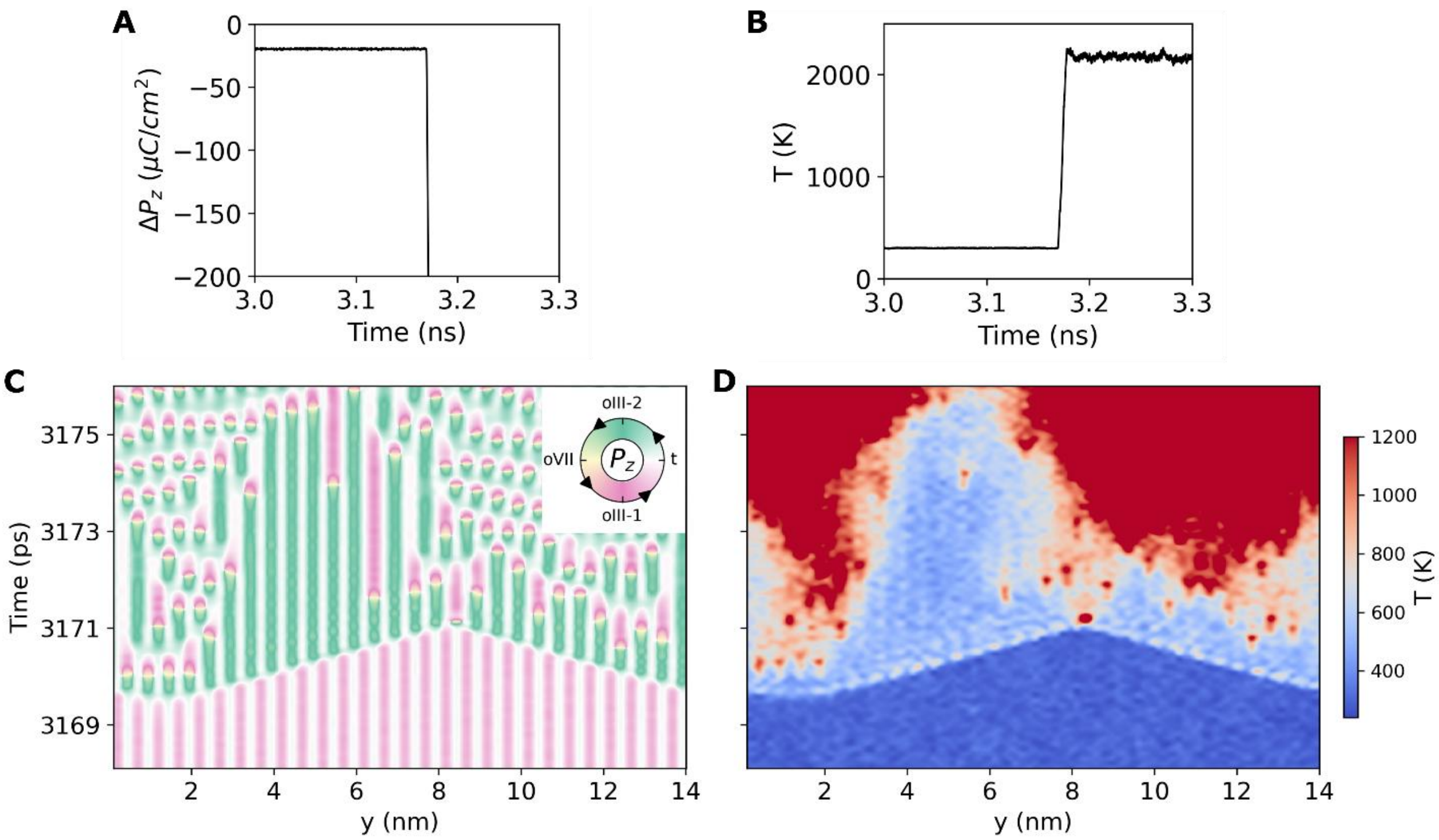

**Fig. 3. Type-A switching kinetics.** Results for an applied field $\mathcal{E} =$ -12.55 MV/cm. (**A**) Polarization ($\Delta P_z$) with respect to the unperturbed oIII-1 state as a function of time. (**B**) Global temperature (T) as a function of time. (**C**) Time evolution of the *Y*-resolved local polarization ($P_z$) averaged over the *XZ* plane. The counterclockwise-pointing arrows in the circular colorbar (top right corner) indicate the sense of the change of $P_z$ induced by negative fields. (**D**) Time evolution of the *Y*-resolved temperature averaged over the *XZ* plane.

Notably, the initial heating prior to the avalanche exhibits important qualitative and quantitative differences depending on whether the first event is type-A or type-B. This is emphasized in Fig. 4 (A), where we compare the local temperature profiles corresponding to both cases in the moments immediately after switching begins and before avalanche is triggered. On the one hand, the localized temperature surge accompanying type-B switching constitutes an excess of ≈626 K at its highest intensity (t=566.8 ps). Nonetheless, because the rest of the simulated supercell remains at ≈300K, the localized spike represents a modest ≈20 K increase in the global temperature (Fig. 4(B)). By contrast, the type-A event yields a more modest local excess temperature, reaching a maximum increase ≈1.7 times smaller than that observed in the type-B event (Fig. 4(A)). However, the heated region is broader and grows rapidly as new channels are switched, so that the global temperature exhibits a much faster increase than in the type-B case (Fig. 4(B)).

This behavior can be quantitatively understood by considering that the electric enthalpy $E = K + V - \Omega \cdot \mathcal{E} \cdot P_z$ (where $K$, $V$ and $\Omega$ are the kinetic energy, potential energy and volume per cell, respectively) must be approximately conserved in timescales that are too short for the heat exchange with the environment to be significant. (Here we ignore entropic effects as they are harder to quantify and not expected to significantly influence the dynamics during the first moments after the initial switching is triggered.) Energy changes during switching should thus follow the relation $\Delta K = \frac{3N}{2} k_B \Delta T \approx -\Delta V + \Omega \cdot \mathcal{E} \cdot \Delta P_z$, where $N$ and $k_B$ are the number of atoms in the cell and Boltzmann's constant, respectively. This is confirmed in Fig. 4(C), where the computed kinetic energy variation (black and red circles) matches our estimate (black and red squares).

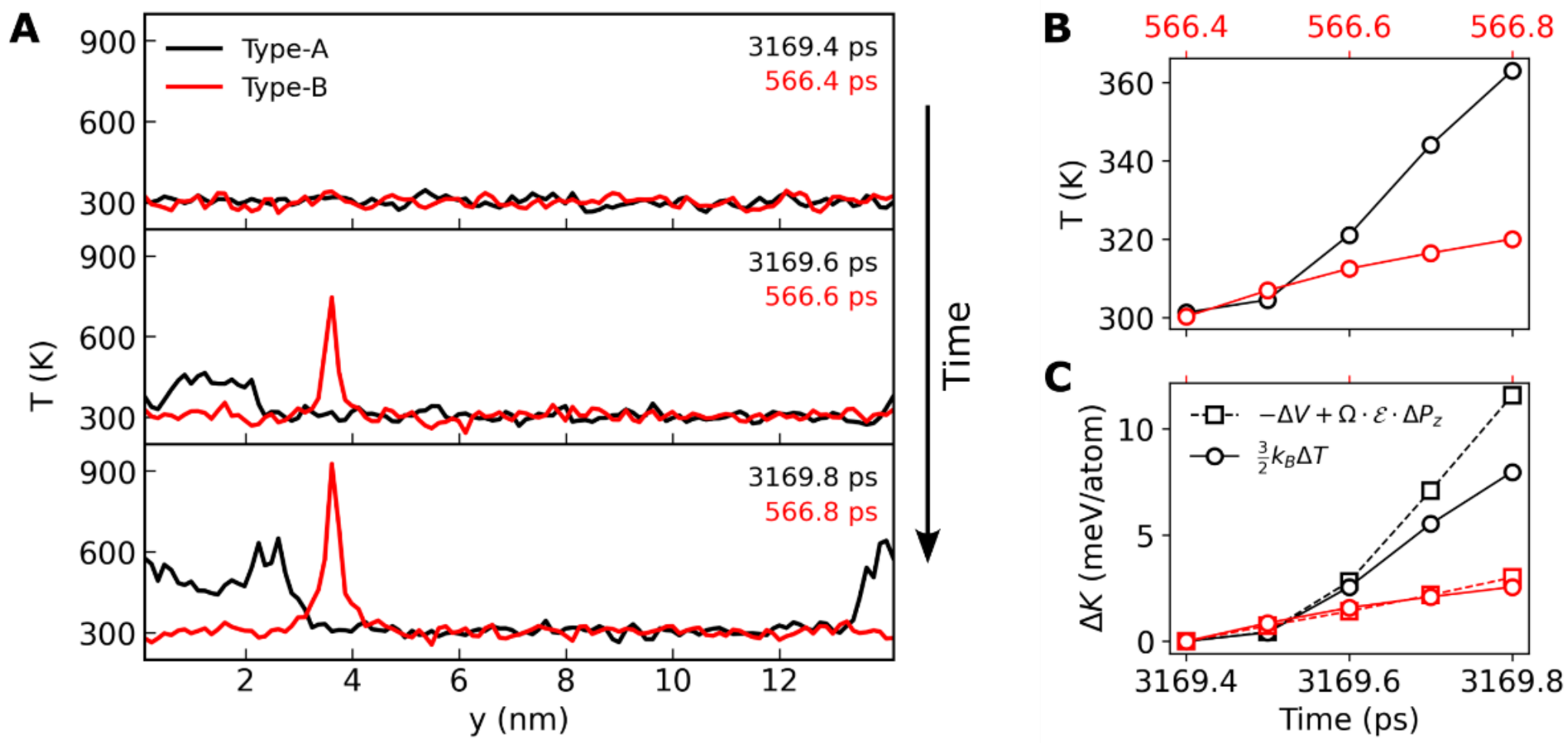

**Fig. 4. Comparison between type-A and type-B switching kinetics.** (**A**) $Y$-resolved temperature (T), averaged on the $XZ$ plane, at different times immediately after type-A (black lines) and type-B (red lines) switching begin. The vertical downwards arrow indicates the sense of increasing time. (**B**) Global temperature during type-A (black circles) and type-B (red circles) switching as a function of time. (**C**) Kinetic energy change ($\Delta K$) as a function of time immediately after the activation of type-A (black) and type-B (red) switching as extracted from the total temperature variation ($\Delta T$) (circles) and the conservation of the electric enthalpy ($E$) (squares). $\Delta K$ is computed with respect to the first timestep in panel (A), i.e. t=3169.4 ps for type-A and t=566.4 ps for type-B. The fields employed for type-A and type-B switching are the same as in Fig. 3 and Fig. 2, $\mathcal{E} = -12.55$ MV/cm and $\mathcal{E} = 14.5$ MV/cm, respectively.

Having established this approximate way to obtain the temperature increase $\Delta T$ from potential ($-\Delta V$) and electric ($\Omega \cdot \mathcal{E} \cdot \Delta P_z$) energy changes, we can gain further insight into the differences between type-A and type-B events. Type-B switching of a single O channel leads to a very stable configuration with 'hard' *Pbca*-like DWs, the change in $\Delta P_z$ saturating to a small value of 4.5 $\mu$C/cm$^2$ (see Figs. 2(A, B)). Despite the strong applied field, $\Delta T$ remains moderate on account of the small induced $\Delta P_z$. This is consistent with, and explains, our observation that an isolated type-B event does not guarantee the activation of O diffusion. Indeed, from the simulations performed in connection to Fig. 1, we find that the time needed to induce O diffusion always exceeds the switching time $\tau$ (fig. S7). By contrast, type-A nucleation leads to a multidomain configuration with 'softer' *P4₂/nmc*-like DWs. Such a configuration is destabilized by the strong applied field, making the monodomain oIII-2 state the next available minimum of the free energy at 300K. This results in the immediate migration of the DWs and the fast growth of $\Delta P_z$, explaining the unstoppable increase of the global temperature and immediate activation of O diffusion.

Hence, the dramatic temperatures reached upon type-A switching modify the free energy landscape, such that the type-B mechanism becomes active (despite its higher $\mathcal{E}_t$, see Fig. 1(D)) and the avalanche occurs. However, it is not obvious whether that behavior is sustained entirely by the heat generated during the initial moments of the switching process or whether it requires strong fields to remain applied. To address this question, we implement a different simulation protocol whereby a negative pulse-like electric field of $\mathcal{E}_{\text{pulse}} = -12.55$ MV/cm is applied to an oIII-1 cell to create an oIII-2 nucleus via type-A switching. Immediately after nucleation (t=3169.6 in Fig. 3 and Fig. 4), the field strength is lowered to a new value $|\mathcal{E}| < |\mathcal{E}_{\text{pulse}}|$ that is held constant for the rest of the simulation.

Figure 5(A) shows the final average polarization ($\langle \Delta P_z \rangle$) obtained after the system equilibrates under constant $|\mathcal{E}| < |\mathcal{E}_{\text{pulse}}|$. The critical field for the migration of the *P4₂/nmc*-like DWs delimiting the transformed oIII-2 nucleus –and thus, for full polarization switching– is found at $|\mathcal{E}_s| = 3$ MV/cm (vertical green dashed line), as evidenced by the abrupt change in $\langle \Delta P_z \rangle$. Interestingly, for $|\mathcal{E}| < 10$ MV/cm, we do not have sufficient excess temperature to induce diffusive behavior, suggesting that a suitable field pulse might achieve switching without breakdown.

The threshold for O diffusion is found at $|\mathcal{E}_d| = 10$ MV/cm (vertical blue dashed line). Specifically, in the range $10 \leq |\mathcal{E}| \leq 12.3$ MV/cm, type-B events become active during type-A switching, but breakdown is not yet triggered. Then, upon reaching the threshold field $|\mathcal{E}_b| = -12.4$ MV/cm (vertical red dashed line), thermal runaway becomes inevitable, O ions diffuse permanently and breakdown occurs.

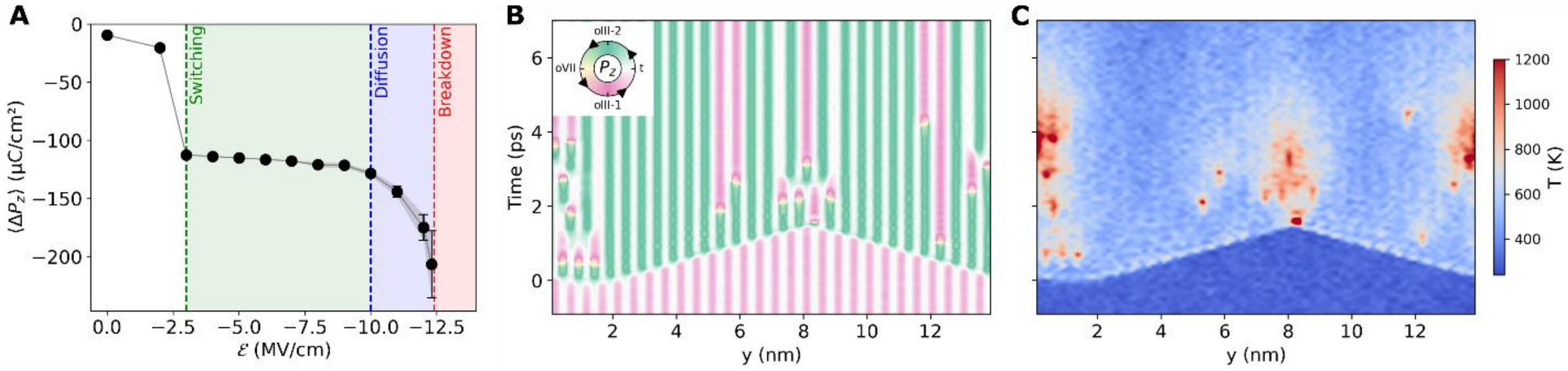

**Fig. 5. Electric field-dependence of switching and diffusion kinetics.** (**A**) Average polarization ($\langle \Delta P_z \rangle$) of the final equilibrium state after switching, calculated with respect to the unperturbed oIII-1 state, as a function of $\mathcal{E}$. In each simulation, the system is thermalized for 30 ps after reaching a stable configuration, and the time-averaged polarization change ($\langle \Delta P_z \rangle$) is extracted over 100 ps after thermalization. We follow this procedure for 5 different simulations per $\mathcal{E}$, from which we extract 5 values of $\langle \Delta P_z \rangle$. The black dots and the error bars at each $\mathcal{E}$ are, respectively, the average and standard deviation of $\langle \Delta P_z \rangle$ over the 5 different simulations. The vertical dashed lines at $\mathcal{E}_s = -3$ (green), $\mathcal{E}_d = -10$ (blue) and $\mathcal{E}_b = -12.4$ MV/cm (red) indicate the thresholds for switching, diffusion and breakdown, respectively. (**B**) Time evolution of the *Y*-resolved local polarization ($P_z$) averaged over the *XZ* plane for a representative case at $\mathcal{E} = -12.3$ MV/cm. At t<0, a monodomain oIII-1 state is subject to a pulse of $\mathcal{E}_{\text{pulse}} = -12.55$ MV/cm. When the oIII-2 domain nucleates (t=0) the field is lowered to $\mathcal{E} = -12.3$ MV/cm. After transient O diffusion, a polydomain state is stabilized, exhibiting multiple oIII-1 and oIII-2 regions delimited by *P4₂/nmc*-like (two adjacent white channels) and *Pbca*-like (a single white channel between pink and green channels) DWs. (**C**) Corresponding time evolution of the *Y*-resolved temperature (T) averaged over the *XZ* plane.

The intermediate regime $10 \leq |\mathcal{E}| \leq 12.3$ MV/cm deserves attention. Here, as the field increases, we observe what we may call *transient avalanches* of subsequent type-B and type-A events (Fig. 5(B) and movie S4). Interestingly, this is accompanied by a large non-linear field dependence of the equilibrium polarization $\langle \Delta P_z \rangle$ (Fig. 5(A)). However, after the transient O flow, the excess heat has enough time to dissipate (Fig. 5(C)) and the system equilibrates to polydomain states characterized by multiple oIII-1 and oIII-2 regions separated by $P4_2/nmc$- and *Pbca*-like DWs (Fig. 5(B) and fig. S8). Remarkably, such polydomain states exhibit high-energy local configurations (see e.g. $P4_2/nmc$-like DWs separating two oIII-2 domains). We find that, upon reduction of the applied field, such local configurations typically transform into lower energy states (fig. S9). Such a field-driven discontinuous transformation suggests an avenue for storing part of the excess kinetic energy in the form of electric potential, which can then be recovered on demand (fig. S10). This evokes energy storage in antiferroelectrics (*29*).

In summary, we have shown that intrinsic ferroelectric switching leads to oxygen diffusion in hafnia ferroelectrics. Our simulations show that intrinsic switching fields are large enough to induce electric breakdown due to a thermal runaway chain reaction, i.e. an avalanche-like process whereby heating associated to local switching builds up to activate O diffusion. This behavior ultimately arises from the unique features of these fluorite-structured oxides, chiefly their acute polymorphism (which enables multiple non-polar transition states) as well as the large ionic displacements and coercive fields associated to switching. Our results also suggest how breakdown might be avoided by using suitable electric field pulses.

Our observations advance our understanding of common reliability issues in hafnia ferroelectrics (*6–8*). They may also be relevant to applications in filamentary resistive memories (*28*), where the interplay between extrinsic and intrinsic mechanisms is not fully understood. In this context, determining how oxygen vacancies – a dominant *extrinsic* factor – affect the present conclusions is an urgent question for future work.

Our study highlights the importance of realistic temperature treatment in simulations. Conventional techniques, where all the simulated atoms are assumed to exchange heat with an external reservoir, may artificially preclude avalanche-like effects as those studied here.

We suspect our findings will be relevant to a wider class of emerging unconventional ferroelectrics – e.g. some exhibiting very large coercive fields (e.g. $Al_{1-x}Sc_xN$ (*30*)) or a coexistence of switching and ionic diffusion (e.g. $CuInP_2S_6$ (*31*)) – where nonequilibrium kinetics are likely to play a central role. We hope the present work will prompt further studies of such exciting effects and the novel opportunities they may enable.

## Acknowledgments:

**Funding:** Work supported by

Luxembourg National Research Fund grant INTER/NSF/24/18804122/PIEZOHAFNIA (X.D.C., J.Í.-G.)

Luxembourg National Research Fund grant BRIDGES/18421428/SWITCHON (co-funded by Intel Corporation) (I.R.-M., J.Í.-G.)

**Competing interests:** Authors declare that they have no competing interests.

**Data, code, and materials availability:** The machine-learned interatomic potential employed in this study, the corresponding training and validation datasets, and the data that support the findings of this article are publicly available at Zenodo via the following link: https://doi.org/10.5281/zenodo.22826946 (*32*).

# Supplementary Materials for

## Intrinsic switching leads to oxygen diffusion and breakdown in hafnia ferroelectrics

Xabier Diaz de Cerio*, Iñigo Robredo-Magro, Jorge Íñiguez-González*

Corresponding author: Email: xabier.diaz-de-cerio-palacio@list.lu (X.D.C); jorge.iniguez@list.lu (J.Í.-G.)

**Materials and Methods**

Machine-learned interatomic potential (MLIP)

All calculations were performed employing a neuroevolution potential (NEP) incorporating explicit treatment of electrostatics (qNEP) (*33*, *34*) as implemented in the GPUMD package (*35*). qNEP is a highly efficient neural network-based potential in which partial ionic charges are latent features of the model that are implicitly learned by fitting the sum of the electrostatic energy (calculated using the latent Ewald summation) and a short-range NEP to the total energy and forces (*34*).

The qNEP model was trained using a first-principles dataset composed of 2,906 atomic structures (see next section for details about first-principles simulations). The atomic structures in the dataset were collected from MD simulations using the on-the-fly training algorithm implemented in the Vienna *Ab-initio* Simulation Package (VASP), which enables an efficient sampling of the configuration space explored at selected thermodynamic conditions (*36*). In particular, we considered 24- and 96-atom hafnia supercells ($\sqrt{2} \times \sqrt{2} \times \sqrt{2}$ and $2 \times 2 \times 2$ repetitions of the 12-atom unit-cell, respectively) of the $P2_1/c$ (m), $Pca2_1$ (oIII) and $P4_2/nmc$ (t) polymorphs at temperatures between 10 and 300 K and zero-pressure conditions. In addition, the dataset also includes configurations induced from applying external electric fields of different magnitudes along the three lattice directions of the oIII polymorph at 300 K. These include equilibrium configurations of the *Pbca* (oI), $Pca2_1$ (oIII), $Pmn2_1$ (oIV) polymorphs, as well as transitions states emerging during ferroelectric switching and O diffusion processes, such as the *Pbcm* (oVII). In order to ensure that the sign of the partial ionic charges was correctly captured, we additionally included the energy, forces, stresses and Born effective charges (BECs) of the fully-relaxed 12-atom $Pca2_1$ (oIII) unit-cell.

The constructed qNEP model includes real- and reciprocal-space contributions to the latent Ewald summation. The radial and angular cutoffs in the short-range part were set to 6 Å and 4 Å, respectively, while the number of radial and angular descriptor components were set to 8 and 6, respectively. Both the radial and angular descriptor functions were each constructed using 8 basis functions. With respect to the angular descriptors, a maximum of 4 three-body descriptors were used. In the loss function used to optimize the model, the loss terms associated with the energies and forces were set to 1, while those associated to the virials and the total charge were set to 0.1. The term corresponding to the BEC was set to 0.02. The model was trained against the entire training set during every iteration of the optimization procedure (i.e. a single batch with size equal to the number of structures in the dataset was employed). The number of neurons in the hidden layer in the neural network was set to 40. Every other parameter was set to the default value in version 4.8 of the GPUMD package.

Figure S1 shows the performance of the model on the training dataset compared to the target density functional theory (DFT) results. The corresponding root mean square errors (RMSE) for the energies, forces and stresses are 0.88 meV/atom, 0.049 eV/Å and 0.177 GPa, respectively. We also tested the model on a validation set formed of 1,800 structures extracted every 100 time steps of the on-the-fly MD runs performed to generate the training set. The performance of the model on this validation set compared to DFT is shown in Fig. S2. The corresponding RMSEs for the energies, forces and stresses are 0.73 meV/atom, 0.045 eV/Å and 0.164 GPa, respectively.

The developed qNEP model, together with the corresponding first-principles training and validation datasets have been made publicly available in a Zenodo repository (*32*). A more complete assessment of the field-driven protocol employed for the exploration of the configuration space and generation of the training data will be reported elsewhere.

Density functional theory simulations

Single-point DFT calculations of the energies, forces and stresses corresponding to each reference configuration in the MLIP training dataset were performed using VASP (*37*, *38*). The exchange-correlation energy was treated within the general gradient approximation (GGA) using the PBEsol functional (*39*). Core electrons were described using the projector-augmented wave method (*40*), while the following valence states were considered explicitly: *2s 2p* for O, and *5p 6s 5d* for Hf. A plane-wave cutoff energy of 500 eV was employed. Brillouin zone sampling for 12-, 24- and 96-atom cells was performed using $4 \times 4 \times 4$, $3 \times 3 \times 3$ and $2 \times 2 \times 2$ *k*-point Monkhorst-Pack grids, respectively.

MD simulations

MD simulations at constant homogeneous strain and temperature (NVT ensemble) conditions were performed using the GPUMD package (*35*, *41*). A simulation time-step of 1 fs was chosen. The strain state was fixed to the equilibrium strain of the oIII polymorph at 300 K and zero electric field. To determine the equilibrium strains at 300 K, a 6,048-atom supercell (6x14x6 unit-cell repetitions) in the oIII phase was heated from 0 to 300 K at a rate of 0.1 K/ps at zero-pressure conditions using a Nosé-Hoover chain thermostat and a Parrinello-Rahman barostat (*42*). After reaching 300 K, the system was thermalized for 400 ps and the strains were averaged over the next 200 ps in the isothermal-isobaric ensemble. Before performing any production simulation, homogeneous strains were fixed to their average equilibrium values, and the system was thermalized for 400 ps at a constant temperature of 300 K (NVT ensemble). We refer to the Supplementary Text and Fig. S5 for a justification of our choice of elastic boundary conditions.

In production simulations, temperature was controlled using stochastic boundary conditions (SBC) (*22*, *23*). In particular, only 4 consecutive horizontal (XY) Hf layers were coupled to a Langevin thermostat at 300 K, while the rest of atoms (including all O ions) followed Newtonian dynamics. The friction coefficient of the Langevin atoms was set to 100 ps$^{-1}$. Using this method, only Langevin Hf atoms are in contact with a heat reservoir, mimicking a thin film that exchanges energy with the environment only through its surfaces or interfaces. A schematic representation of the way in which the simulation supercell was coupled to the thermostat is provided in Fig. S3. Importantly, this approach is suitable for simulating equilibrium properties, as it accurately reproduces ionic dynamics according to the canonical (NVT) ensemble (*23*). Moreover, SBC account for a realistic treatment of energy dissipation during nonequilibrium processes (*22*). In fact, we find that our results can be drastically affected by the use of conventional thermostat techniques, as those can provide excessively damped dynamics that mask the actual physics of the field-driven transformations in the system (see Fig. S4).

The external electric field was applied using the force method, whereby the force induced by a field $\vec{\mathcal{E}}$ on ion *i* is given by $\vec{F}_i = q_i \times \vec{\mathcal{E}}$, where $q_i$, the formal charge of ion *i*, is +4e for Hf and -2e for O. The polarization was calculated as $\Delta\vec{P} = \sum_i q_i \, \Delta\vec{r}_i$, where $\Delta\vec{r}_i$ is the displacement of ion *i* with respect to the oIII-1 state at zero-field.

Visualization

Structural models were generated using VESTA (*43*), while OVITO (*44*) was used to visualize MD trajectories and generate the animations.

**Supplementary Text**

Influence of the thermostat in the ionic dynamics

In order to illustrate the influence of the thermostat employed in the MD simulations, we repeat the simulation performed in Fig. 1(b) and we compare the system behavior when we use SBC (same as the calculation made in connection to Fig. 1(b)) and when we apply the conventional procedure in which all the atoms in the simulation supercell are coupled to the thermostat. In particular, the simulation consists of a 6,148-atom supercell in the oIII-1 state with an applied negative external field along the Z-direction, which is activated at t=200 ps and whose magnitude increases linearly in time at a rate of 5 MV/cm per ns. In the case where all the atoms are coupled to the thermostat, we further compare the results obtained from using Nosé-Hoover (*42*), stochastic velocity rescaling (*45*) and Langevin thermostats. The obtained results are summarized in Fig. S4, which shows the evolution of the global system temperature with time in each case.

The global temperature increase observed in all cases between 2.76 and 2.78 ns indicates the activation of type-A switching. All cases show excellent agreement in the predicted threshold fields (between 12.81 MV/cm and 12.86 MV/cm), as expected from the fact that the latter is an equilibrium property and that all the employed thermostats are suitable for simulating equilibrium properties in the NVT ensemble.

However, the nonequilibrium evolution of the system following the activation of type-A switching differs drastically depending on the thermostat. The employed Nosé-Hoover thermostat shows the closest qualitative agreement with the SBC method employed in this work. In particular, type-A switching immediately activates O diffusion and leads to the breakdown of the system due to thermal runaway. In the case of stochastic velocity rescaling, O diffusion is also activated after switching, but the produced kinetic energy increase is more efficiently dissipated by the thermostat. Hence, O diffusion is lower and does not appear in the whole supercell at the same time. Finally, the case of the Langevin thermostat represents an extreme limit, in which the local heat produced during type-A switching is rapidly dissipated. Consequently, the induced temperature increase remains moderate and decreases immediately, and type-A switching does not lead to any O diffusion event, corresponding to a transition from the homogeneous oIII-1 to the homogeneous oIII-2 state.

The role of strain: fixed vs free elastic boundary conditions

In the present work, we have considered the homogeneous strain of the system to be fixed (NVT conditions). This choice is justified by different factors. First, to the best of our knowledge there is currently no implementation of SBC within MD simulations with free strains (NPT conditions) where we can use our qNEP model. Second, hafnia exhibits a low piezoelectric response (*46*). As such, we do not expect the applied threshold fields to induce lattice deformations that are sufficient to qualitatively change the ionic dynamics. Third, thin films in experiments are typically epitaxially clamped by the substrate. If the films are thin enough, we could expect the strain clamping to extend to their whole thickness. Thus, we consider the case with fixed strains to be experimentally relevant.

Nevertheless, one could wonder how macroscopic changes in strain affect our conclusions. To make an estimate, we have repeated the simulations performed in connection to Fig. 1(b) by allowing the strains to relax freely under zero-stress conditions during the MD run and we have compared the results to those obtained fixing the strains (same as Fig. 1(b)). Noting that we cannot perform free strain simulations with SBC, we have employed a Nosé-Hoover chain thermostat and a Parrinello-Rahman barostat (*42*) to perform the NPT runs. The results are summarized in Fig. S5(B), where we show the change of polarization ($\Delta P_z$) induced by an electric field ($\mathcal{E}$) with magnitude increasing linearly in time. We observe that unclamping the cell and letting the strains relax leads to a slight dielectric softening of the material for both positive and negative fields, as demonstrated by the somewhat faster evolution of $\Delta P_z$ with the field in those cases. Consequently, the threshold field for type-A switching in the unclamped case is reduced by 0.8 MV/cm (i.e., about 6%) compared to the clamped situation. In the case of type-B switching, the threshold field does not exhibit any significant variation.

These small changes are not expected to affect qualitatively our conclusions. Notably, despite the fact that the Nosé-Hoover chain thermostat damps the ionic dynamics compared to SBC (see fig. S4), Fig. S5(B) shows that high ionic diffusion and temperature increase are also induced upon switching when the strains are free to relax. Therefore, we do not expect the rest of simulations performed in this work to be significantly affected by our choice of elastic boundary conditions.

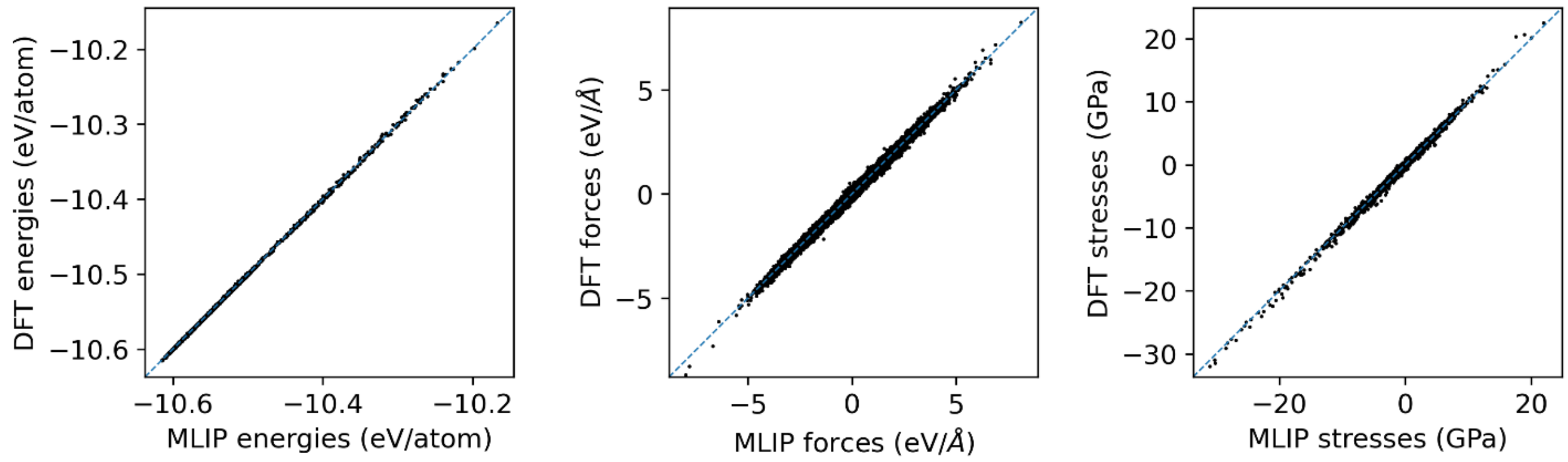

**Fig. S1.**

Comparison between energies, forces and stresses predicted by our qNEP MLIP and DFT for all the configurations in the training dataset.

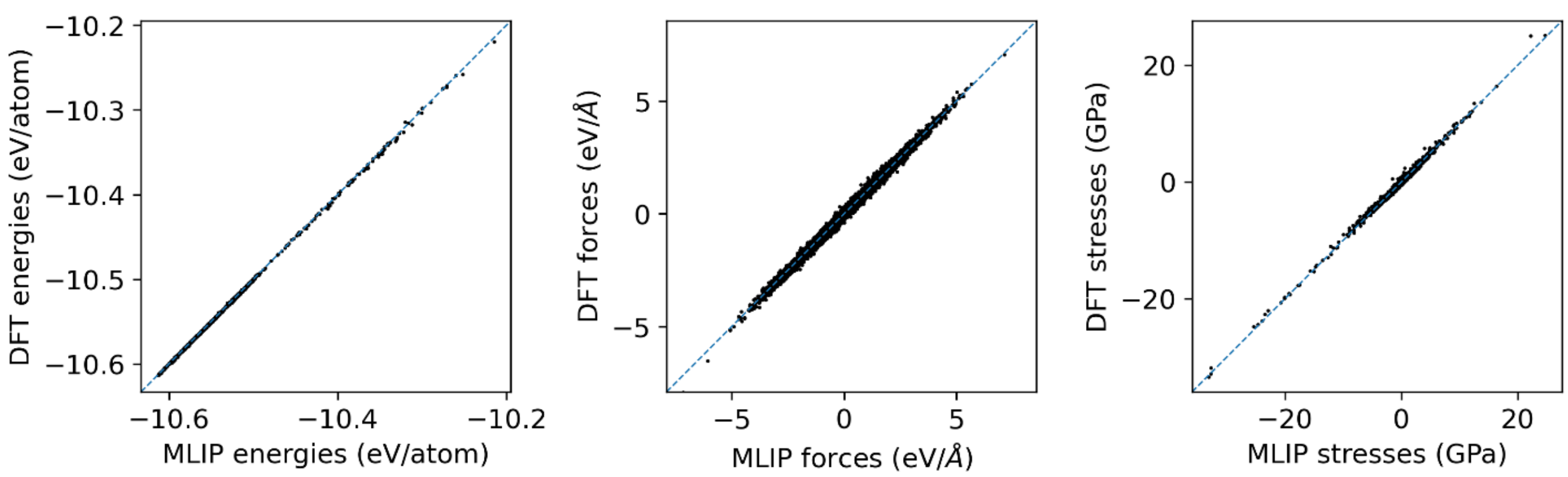

**Fig. S2.**

Comparison between energies, forces and stresses predicted by our qNEP MLIP and DFT for the configurations in the validation set.

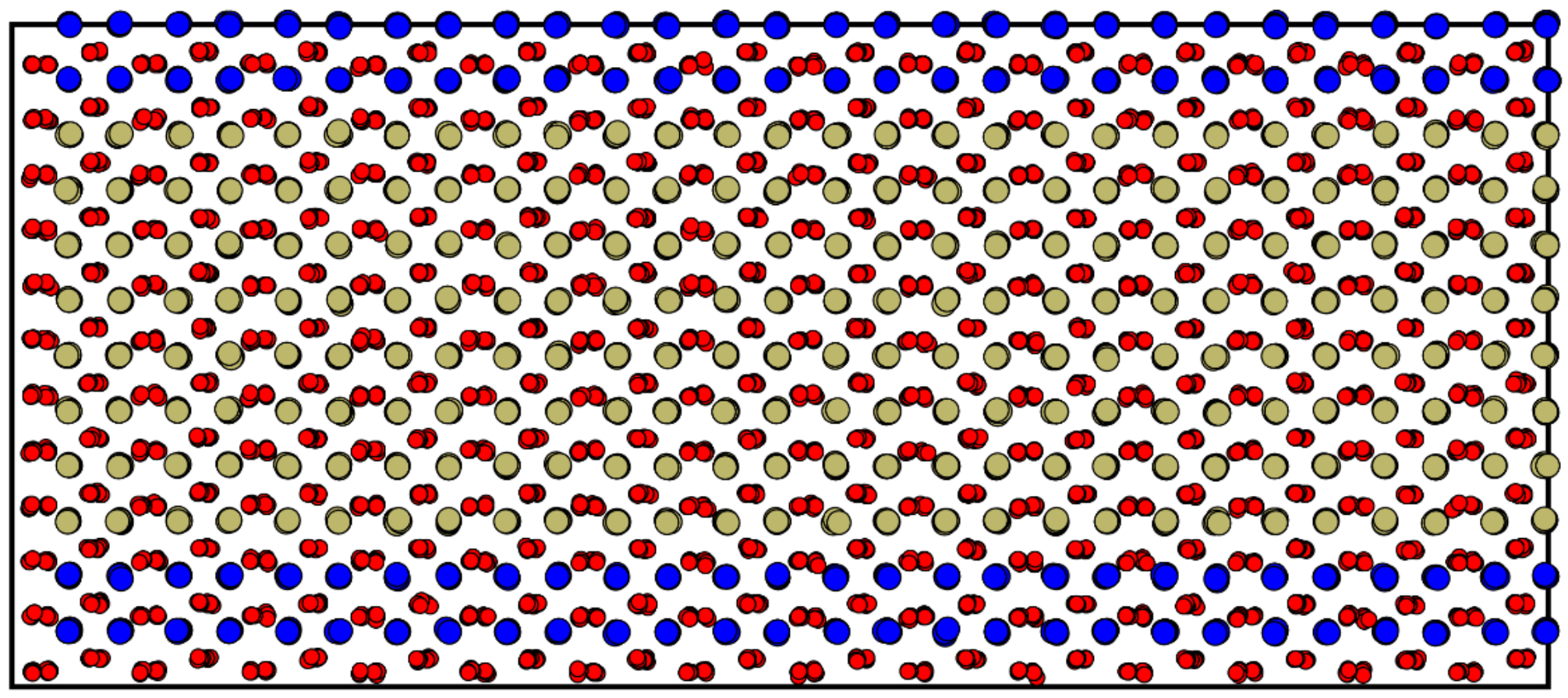

**Fig. S3.**

Snapshot of a 6,048-atom supercell (6x14x6 unit-cell repetitions) in the oIII-1 state at 300 K. The Hf ions coupled to the Langevin thermostat are represented in blue, while Newtonian Hf and O ions are khaki and red, respectively.

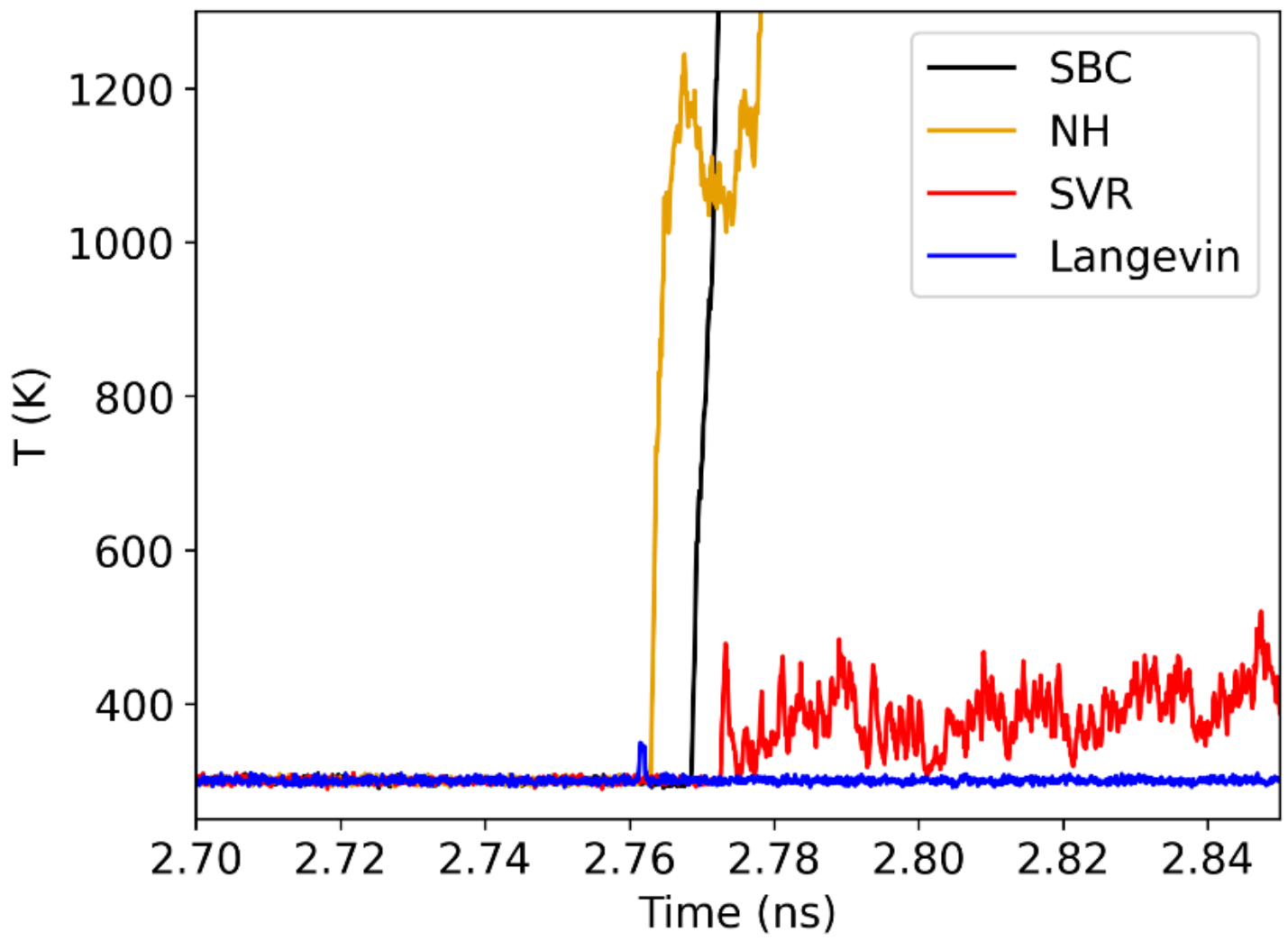

**Fig. S4.**

Evolution of the global system temperature (T) as a function of time in the case where a linearly increasing electric field is applied to a 6,148-atom oIII-1 supercell, as in Fig. 1. Each color corresponds to a different thermostat employed for the same simulation: SBC (black), Nosé-Hoover (NH, yellow), stochastic velocity rescaling (SVR, red), and Langevin (blue). For the parameters of the NH, SVR and Langevin thermostats we employed the default values in version 5.0 of the GPUMD package. In the case of SBC, the friction coefficient in the Langevin atoms was chosen to 100 $ps^{-1}$.

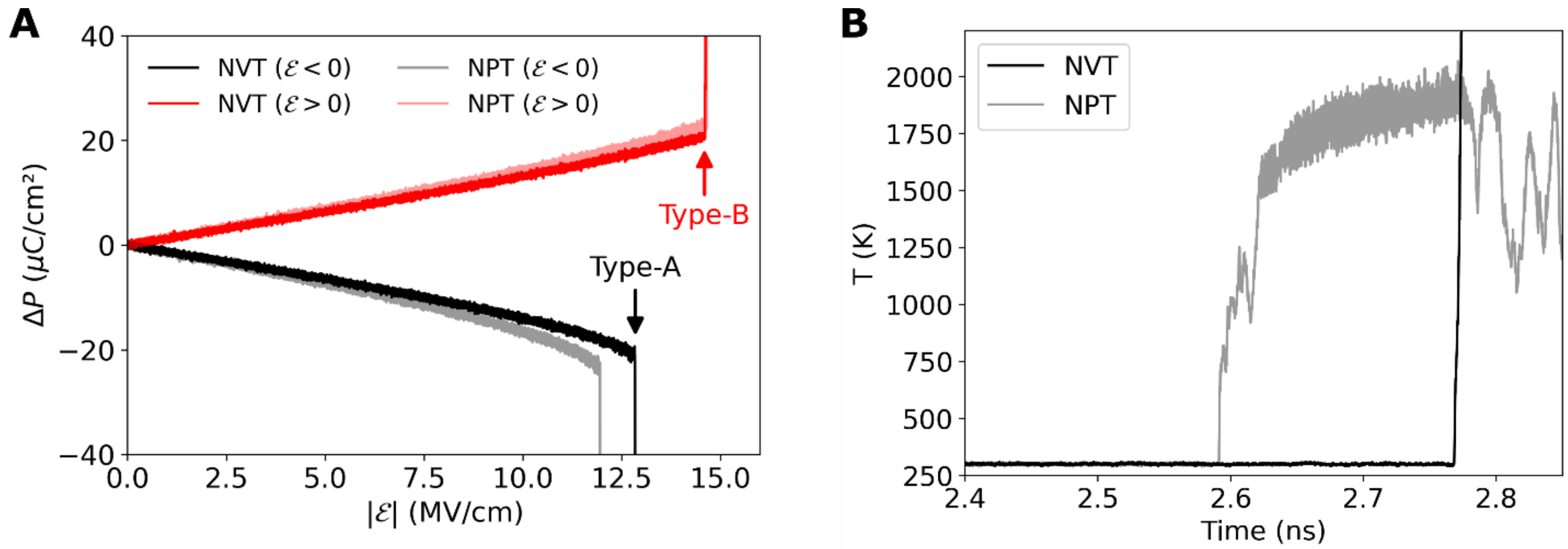

**Fig. S5.**

(A) Induced change in polarization ($\Delta P_z$), calculated with respect to the unperturbed oIII-1 state, as a function of external electric field strength ($|\mathcal{E}|$) for $\mathcal{E} < 0$ (black) and $\mathcal{E} > 0$ (red). Dark (faint) lines correspond to simulations where the strains were fixed (free to relax). $|\mathcal{E}|$ is linearly increased over time at a constant rate of 5 MV/cm per ns. Regardless of whether the strains are fixed or free to relax, $\mathcal{E} < 0$ and $\mathcal{E} > 0$ induce type-A and type-B switching, respectively. Also, in all cases, switching is followed by permanent oxygen diffusion and a giant increase in temperature. (B) Evolution of the global system temperature (T) as a function of time for the cases with $\mathcal{E} < 0$ in panel A. We have zoomed the time range where switching is activated.

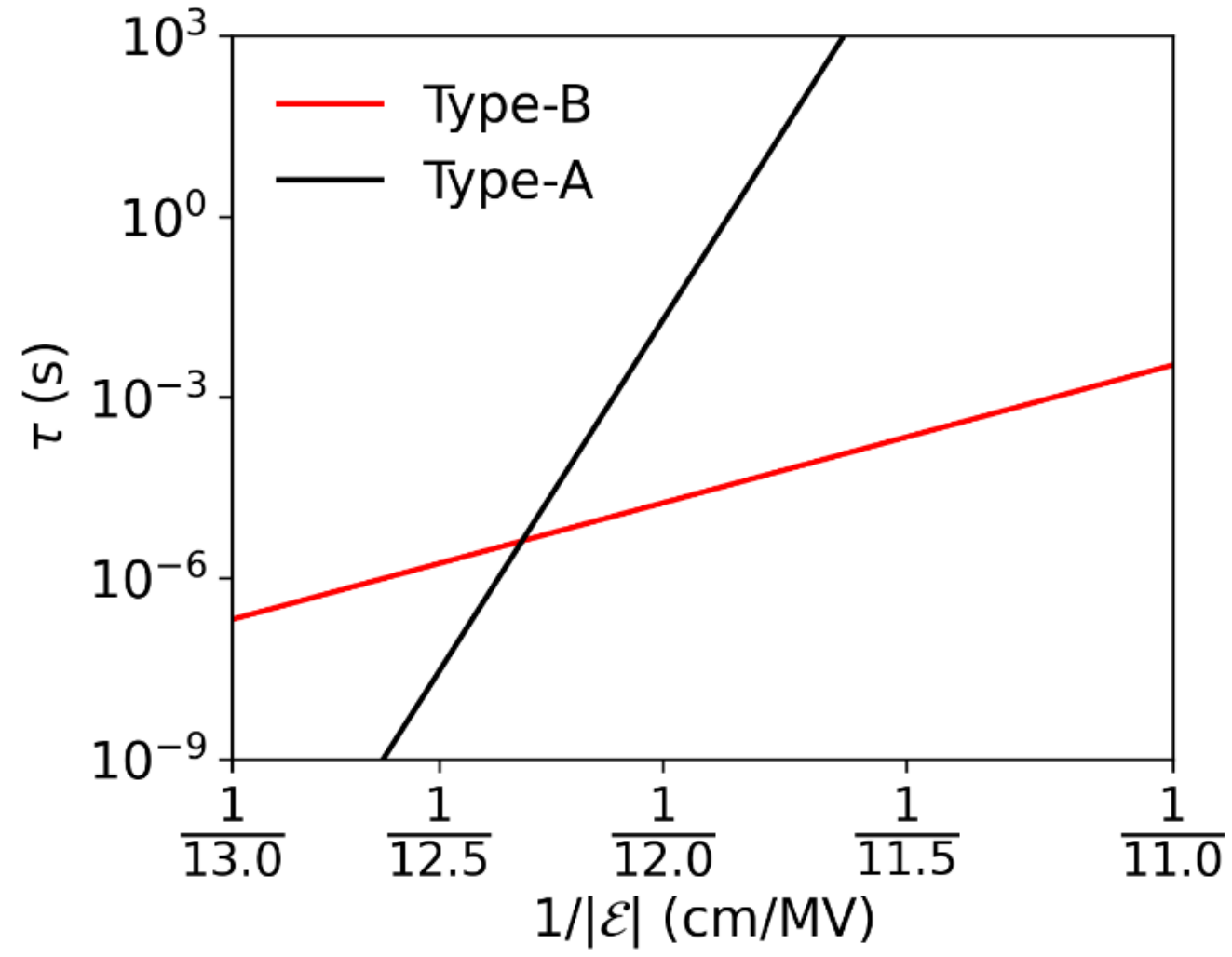

**Fig. S6.**

Extrapolation of the Merz-like behavior of switching activation times ($\tau$) to lower fields $\mathcal{E}$. The fit of the $\tau$ vs. $1/|\mathcal{E}|$ relation to the empirical Merz's law for type-A and type-B switching is given by the black and red lines, respectively. According to the extrapolation of the Merz-like behavior, type-A and type-B switching become equally likely for $|\mathcal{E}|$ ≈12.3 MV/cm, with $\tau \approx 4\ \mu s$.

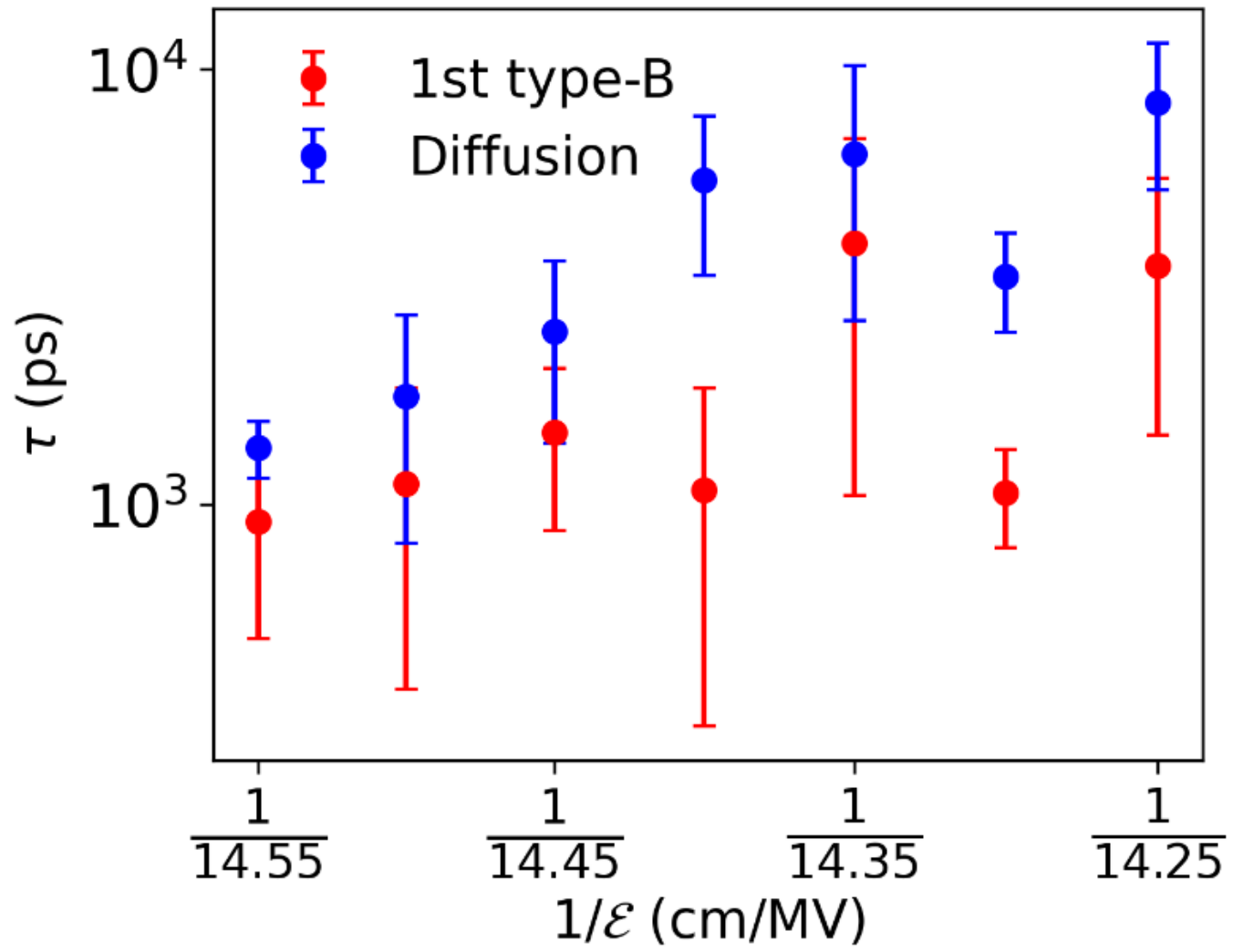

**Fig. S7.**

Type-B switching and diffusion activation times as a function of $1/\mathcal{E}$. The times needed for the first event and for the event triggering permanent O diffusion are indicated by the red and blue dots, respectively. The dots and the error bars correspond to the average and standard deviation over the 3 simulations performed for each field. The figure shows that the times needed to activate O diffusion are always higher. This result indicates that several type-B events are needed before the avalanche-like process is triggered and permanent O diffusion is induced. We also find that the avalanche is typically triggered when the type-B event happens close to a previously switched oIII-2 domain (usually featuring at least 2 adjacent switched channels), as this can undergo subsequent type-A switching more easily.

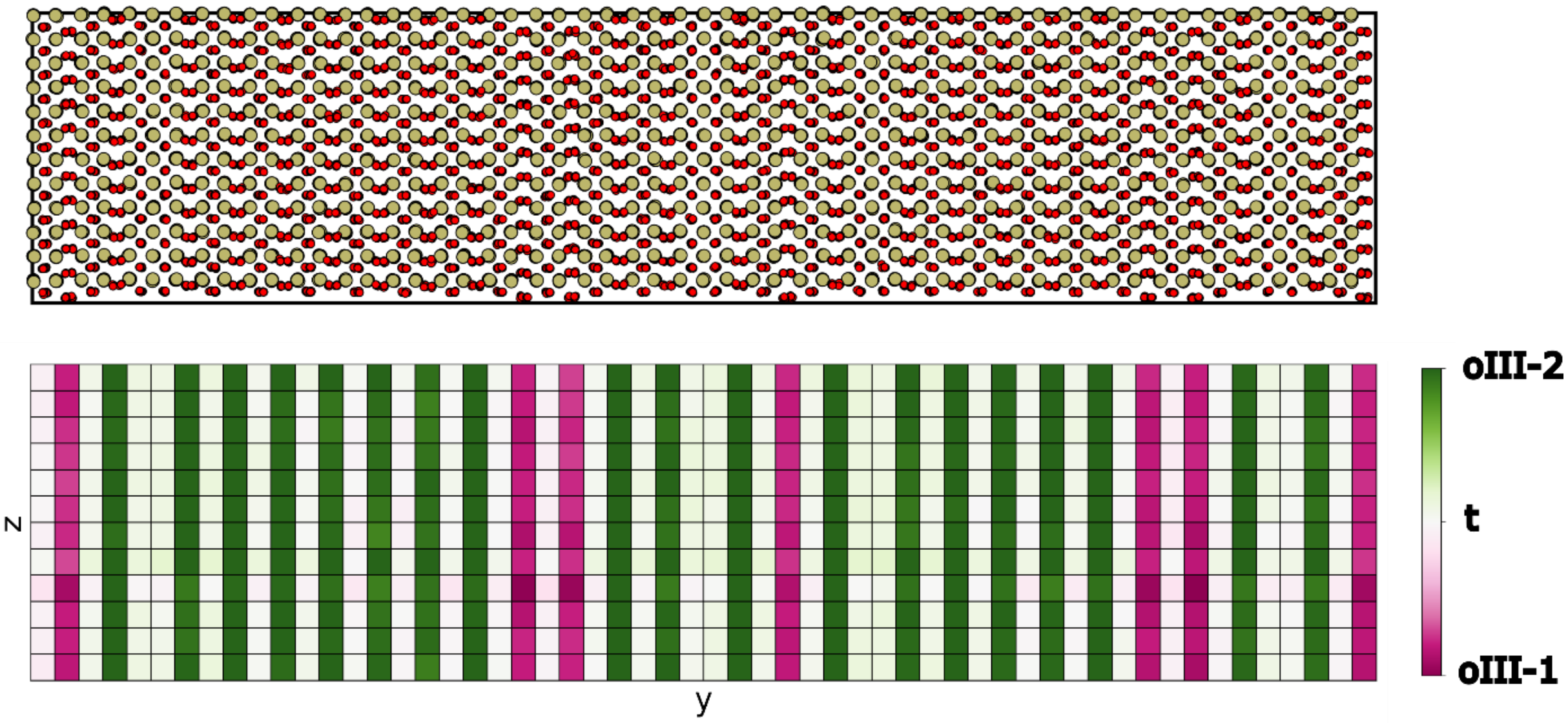

**Fig. S8.**

Top: Atomic structure model of the final equilibrium configuration reached in Fig. 5(b), with $\mathcal{E} = -12.3$ MV/cm. Bottom: schematic representation of the local "$P_z$ state" corresponding to the configuration in top panel. The "$P_z$ state" is calculated for each O atom in the supercell, and then averaged over the *X* axis and resolved in *Y* and *Z*. The represented configuration exhibits *Pbca* (oI)-like DWs separating oIII-1 and oIII-2 domains (a single white channel between pink and green channels), as well as *P4₂/nmc* (t)-like DWs between different oIII-2 domains (two adjacent white channels between two green channels). The latter is a high energy local configuration stabilized by the field (see Figs. S9 and S10).

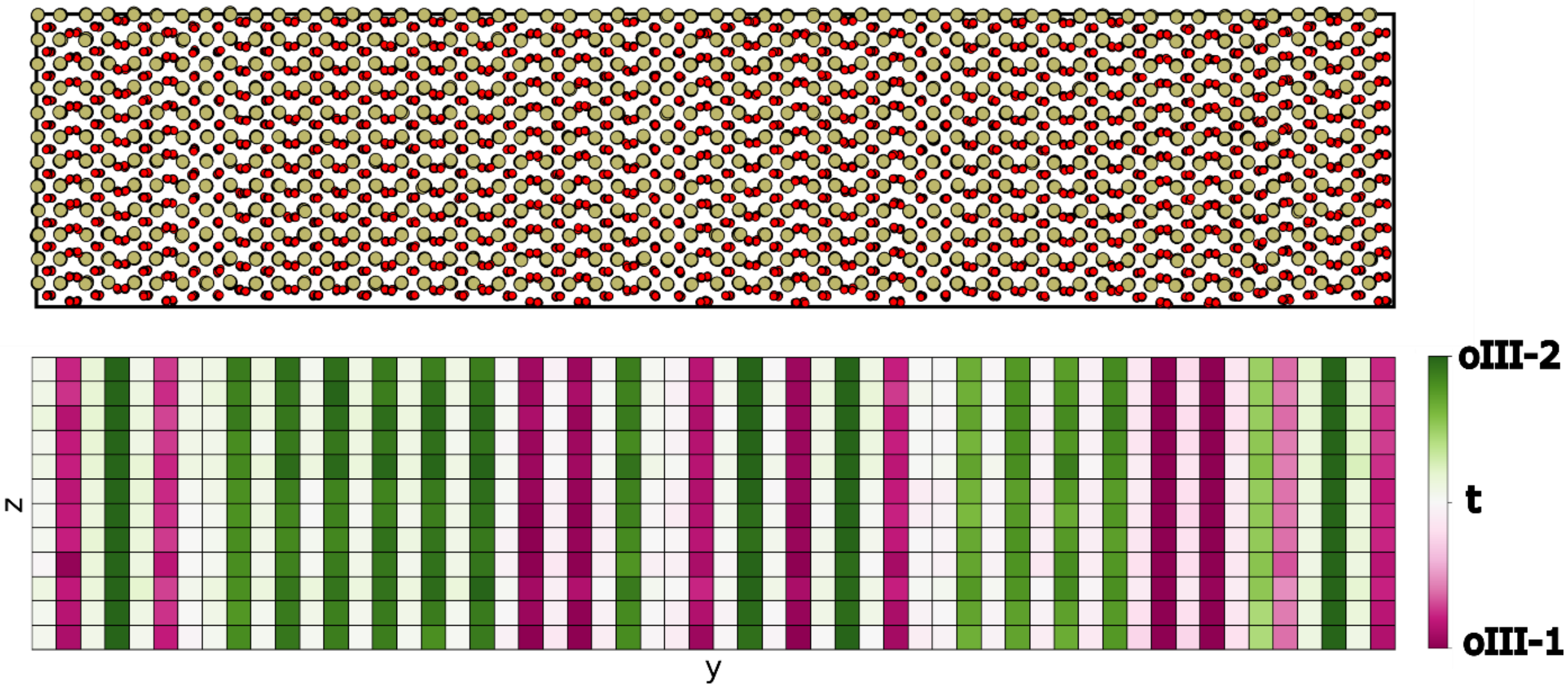

**Fig. S9.**

Top: Atomic structure model of the configuration reached at $\mathcal{E} = 0$ MV/cm after gradually decreasing the field in the configuration shown in Fig. S8. Bottom: schematic representation of the local "$P_z$ state" corresponding to the configuration in the top panel. The *P4₂/nmc* (t)-like DWs between different oIII-2 domains, shown in Fig. S8 for $\mathcal{E} = -12.3$ MV/cm, are destabilized as the field is reduced, eventually resulting in *P4₂/nmc* (t)-like DWs between oIII-1 and oIII-2 domains. In one of the cases the unstable DW transforms locally into a *Pbcn*-like configuration. Notably, these transitions induce discontinuous changes of polarization (see Fig. S10 for further details).

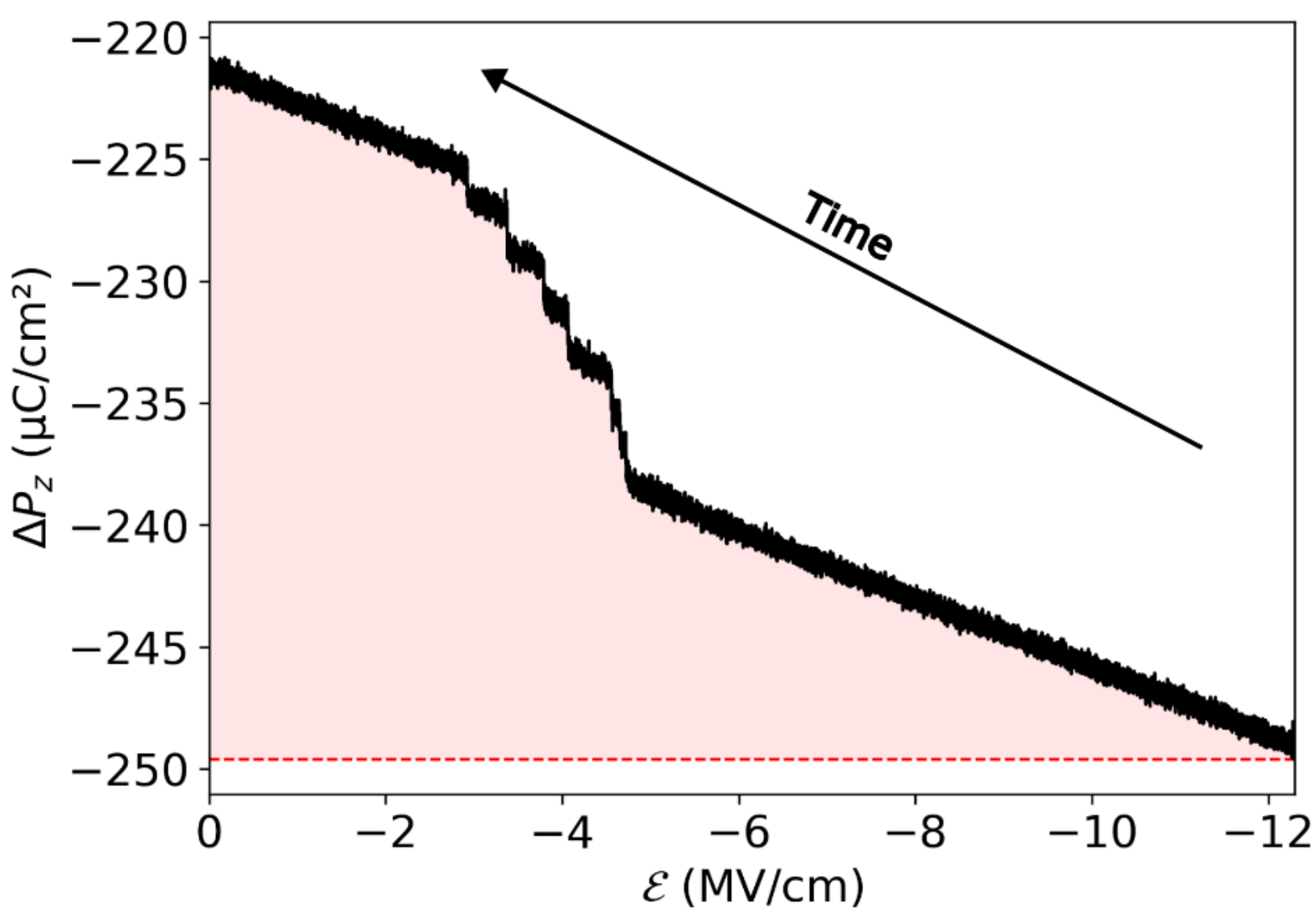

**Fig. S10.**

Polarization ($\Delta P_z$), computed with respect to the unperturbed and homogeneous oIII-1 state, as a function of $\mathcal{E}$ for a system initially in the polydomain state shown in Fig. S8 and with $\mathcal{E} = -12.3$ MV/cm. The field is gradually reduced to 0 MV/cm during the simulation at a constant rate of 5 MV/cm per ns. The arrow indicates the sense of increasing time. As $\mathcal{E}$ decreases some of the local configurations in the initial state are destabilized and transition to lower energy configurations (see Figs. S7 and S8), inducing positive discontinuous changes of polarization. Eventually, at $\mathcal{E} = 0$ MV/cm, the system reaches the configuration shown in Fig. S9. The energy recovered after reducing the field is represented by the red shaded area. The described discontinuous transitions enhance the recovered energy.

**Movie S1.**

Animation of the first type-B switching event shown in the bottom part of Fig. 2(b) ($t \approx 567$ ps). Hf atoms are yellow and O atoms are red.

**Movie S2.**

Animation of the thermal runaway process induced by type-B switching shown in the top part of Fig. 2(b) ($t \geq 1052$ ps). Hf atoms are yellow and O atoms are red.

**Movie S3.**

Animation of the thermal runaway process induced by type-A switching shown in Fig. 3(c) ($t \geq 3169$ ps). Hf atoms are yellow and O atoms are red.

**Movie S4.**

Animation of the transient avalanche shown in Fig. 5(b) for $t>0$. After applying a pulse of $\mathcal{E}_{\text{pulse}} = -12.55$ MV/cm and nucleating an oIII-2 domain via type-A switching, the field is reduced to $\mathcal{E} = -12.3$ MV/cm and left constant for the rest of the simulation. The animation shows the ionic dynamics right after nucleation is triggered and the field is reduced to $\mathcal{E} = -12.3$ MV/cm. After transient O diffusion, the system equilibrates to the final configuration shown in Fig. 5(b) and Fig. S8. Hf atoms are yellow and O atoms are red.